**Wiktor B. Daszczuk**

# DISCRETE EVENT SIMULATION OF PERSONAL RAPID TRANSIT (PRT) SYSTEMS

***Abstract.*** *The article discusses issues related to the construction of the PRT network simulator and the simulation process: the elements of PRT network structure, their representation in the simulator, the simulation process itself, animation, and automation of the experiments. An example of a simulation environment Feniks is described, elaborated within the framework of the Eco-Mobility project.*

## INTRODUCTION

Personal Rapid Transit (also called Automated Transit Network or PodCars) is a modern urban transport system, in which driverless 4-6 person vehicles use a guideway separated from road movement (usually elevated). It performs „door-to-door" transit, without intermediate stops. PRT will possibly support urban transport solution, especially in areas where traditional transport is inconvenient, for example in parks, university campuses, fair areas etc.

In Eco-Mobility project [28] an own concept of PRT network was elaborated, with vehicle model, experimental working prototype in (guideway and vehicles) in ¼ scale. Several aspects of statics and dynamics of PRT vehicles were analyzed [17,18,19,38], and a PRT network simulator Feniks was elaborated. The purpose of the simulator is an analysis of PRT network operation, on coordination and management levels. During the course of the project, several versions of the simulator were developed, the current version is 4.

## 1. PRT SIMULATION MODEL

### 1.1. Overview

A PRT network model is needed for simulation experiments. The model contains:

1. Network topography (guideway graph: nodes, edges, distances, location on the city map, kinds and parameters of stations and capacitors, etc.),
2. A set of vehicles (number of vehicles and their parameters: capacity, maximum velocity, acceleration, deceleration, distance between vehicles in movement),
3. PRT services demand model (time distribution between client appearances, intensity if input stream, dependence on time of day and topography),
4. Passenger behavior model (rules of grouping, service cancellation conditions, ride sharing possibility, target station choice rules, boarding and alighting times),
5. Vehicle control parameters (keeping up algorithm, separation maintaining, priority rules on join intersections, vehicle behavior at a station, turning the movement),
6. Parameters of vehicle set management (including empty vehicles management, dynamic route guidance with regard to congestion in the network, etc. [7,11,42,44,50,53]),
7. Events subject to registration.

A fully defined simulation model is a set of data, determining the effect of the simulator operation during the simulation experiments. In order to increase the statistical reliability of the recorded data, simulation experiments should be repeated for exactly the same set of parameters of the model (an experiment should consist of many replications). To measure a dependence of a given factor on a value of the parameter (for example a dependence of passenger waiting time on a number of vehicles in the network) a number of simulation experiments should be performed (each with several replications) in which one model parameter (e.g., the number of vehicles) changes, while others are fixed. Thus, during the experiments the simulator performs a substantial number of simulation runs, and both the experiments planning and the technique of recording and processing their results play a very important role.

The Eco-Mobility project also covered a more general simulation research, aimed at detecting the more universal quality dependencies (e.g., dependence of passenger waiting time on the general empty vehicle management algorithm parameters [21,22,23]), not connected to the network with specific, given topography. In these studies, to increase the representativeness of the quality of applications, a few simplified network models were used, which corresponded to a hypothetical, possibly a typical layouts hypothetical network PRT: network in a traditional arrangement with the center and the periphery, in the city of streets forming a rectangular grid, the field stretched linearly (along the sea coast for example), etc. These examples of the network play the role of benchmarks [43].

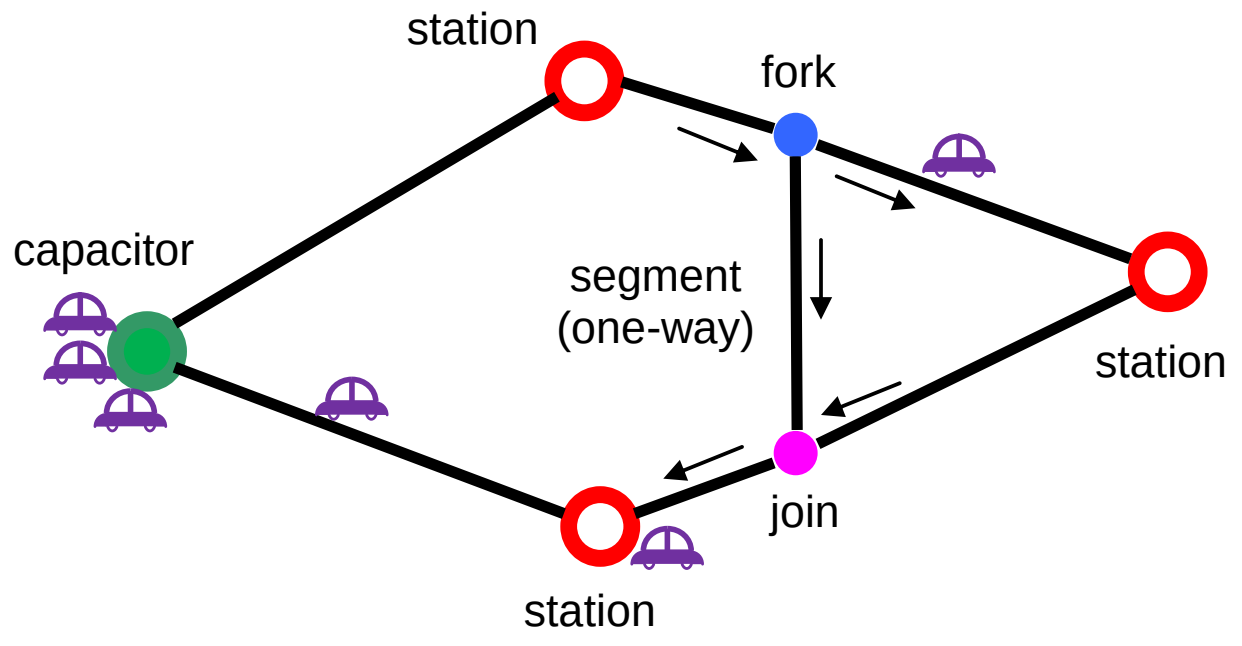

***Fig. 1*** *PRT network graph*

## 1.2. Network Topography

The diagram of PRT guideway can be considered a directed graph consisting of nodes and edges. Nodes are capacitors, stations and intersections. Intersections connect segments of the track: fork (1 → 2) and join (2 → 1). The edges are one-way segments of the guideway. This is illustrated in Fig. 1.

Because the computer simulation is the process by nature discreet, therefore the movement of vehicles on the segments of the track must be digitized. For this purpose, each edge of the graph is logically divided into sectors. The number of sectors in each segment must be integer. Junctions of successive sectors are the points where the movement parameters (the current state of the vehicle) are determined. The parameters are applicable until the end of the next sector, where they will be recalculated.

The length of the sector is the parameter of the simulation, not a part of the network topography. Obviously, the smaller is the length of the sectors, the more accurately the vehicles movement is simulated. But at the same time the cost of the calculations is bigger and the simulation is slower. Therefore the length of the sectors is usually taken to be a compromise, that should be several times (typically 2-3 times) shorter than planned separation distance between vehicles.

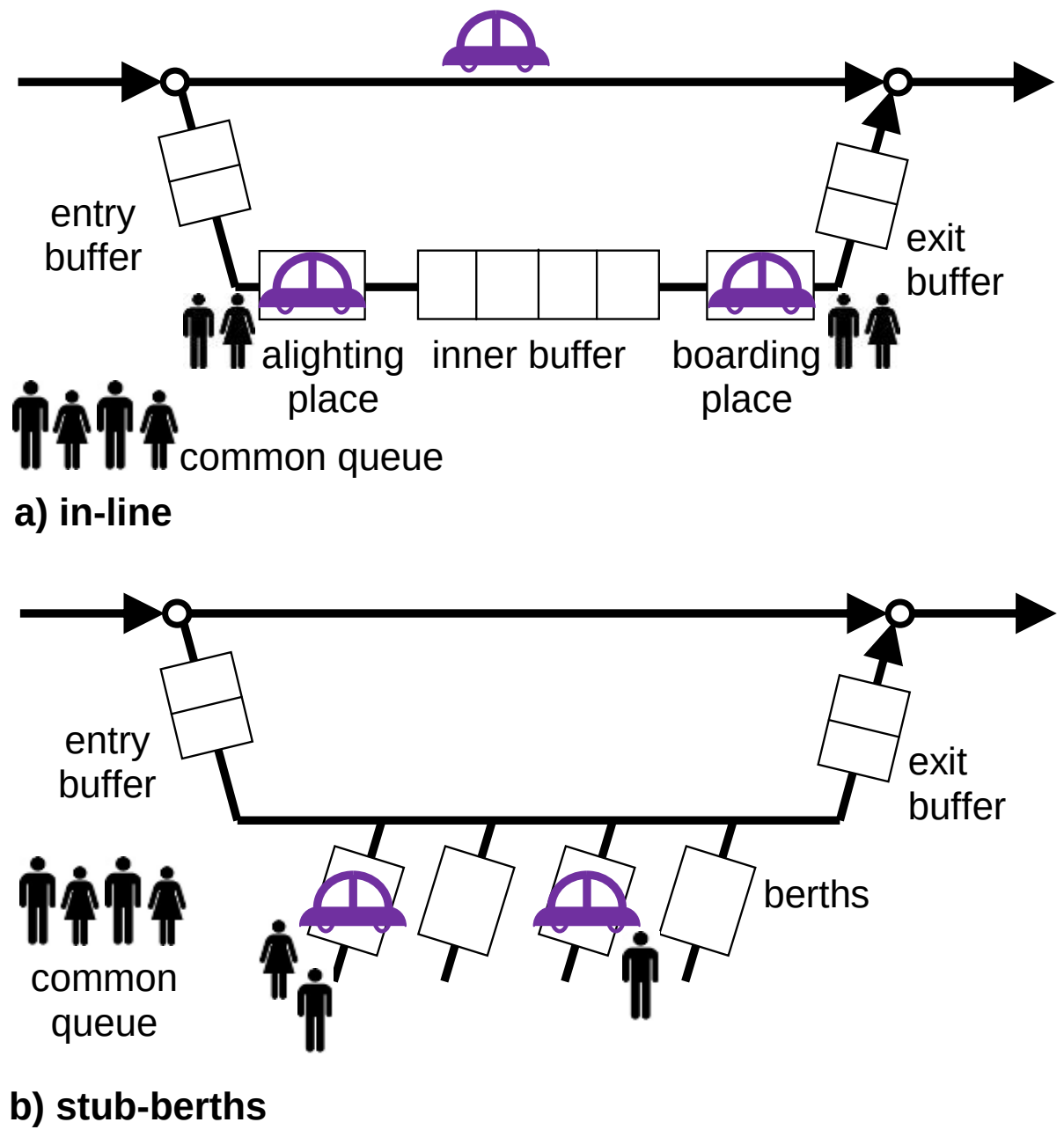

***Fig. 2*** *Station types: a) in-line b) stub-berths*

## 1.3. Stations

A station is a place where the passengers begin and end their trips. The functions of the station determine the occurrence of certain elements of its structure: the passenger queue, the parking places for the vehicles (the berths) and the input and output buffers. The main types of stops are: in-line and stub-berths (they are presented in Fig. 2).

## 1.4. Capacitors

A capacitor (garage, depot) is a source of vehicles, and a place for storage of vehicles which currently have no trips scheduled. A capacitor is a simplified station (no passenger queue, no boarding and alighting places).

A simulator may be built without the capacitors (e.g., Hermes [35]). In this case, the stations are the source of vehicles (there cannot be more vehicles than parking spaces at the stations). The advantage of this approach is that the warm-up period is much shorter before the model reaches an equilibrium state, because vehicles do not have to "move" to the stations after the start of the simulation. The disadvantage is the lack of a possibility of testing real management algorithms, which must take the capacitors into account.

## 1.5. Vehicles

The vehicles are the only objects moving through the network. In the model, some global parameters should be defined for the whole set of vehicles, such as:

- the total number of vehicles,
- the maximum number of passengers in a vehicle,
- maximum speed,
- acceptable acceleration and deceleration,
- acceptable deceleration during friction braking (in emergency conditions)
- the initial placement of vehicles in capacitors and/or stations.

The project Eco-Mobility assumed that vehicles are only of one type, although in general they may be of several types (e.g., passenger and freight, ordinary and emergency, larger and smaller, etc.).

In addition to the data common to all vehicles, there are needed data structures representing individual vehicles during the simulation. In such a structure for an individual vehicle, in addition to its identifier (number) the motion parameters are stored (such as maximum speed, acceleration and deceleration), its current status (e.g., empty or with passengers, during the trip, boarding, alighting, at the capacitor, at the station or on the track, etc.) and information about the current trip: the cardinality of the group of passengers, the origin and the destination of the trip, travel time (riding, boarding and alighting). The structure must also store the information about the current position of the vehicle (or information enough to calculate the position) needed to implement and keep up on the track and the priority rules on the join intersections.

These data structures are created automatically in the simulator during the start of the experiment (and manages them during the simulation). However, the user should be allowed the possibility of defining additional optional parameters to be stored in the data structures of the vehicles (e.g., for recording additional data from the run of the simulation).

## 1.6. Demand Model

A specific PRT feature is to pursue the rips at the request of a group of passengers [7,25], without the "mounting up" of the new passengers during the trip. A whole group of passengers appears at the station at once (usually of size 1-6 people), and perform a common trip to the selected destination station. Therefore, indivisible groups of passengers reside in the queue at the station, rather than single passengers.

Based on the results known from the queuing theory, and the results of many experiments, the rule of demand in a PRT network is a random Poisson stream of customer requests [33,37]. A client is in this context is a group of passengers wishing to travel together in one vehicle, while the event is the appearing of such a group at one of the stations. As the result, the time $T$ between successive appearances (inter-arrival time) is a random variable (continuous) with an exponential distribution, which means that its probability density function is defined as:

$$f(t) = \lambda e^{-\lambda t} \qquad (1)$$

and the average time between the appearances of passenger groups is $E(T) = 1 / \lambda$. Thus, for example, if there is average of 60 groups of passengers per hour, then the intensity of the stream is $\lambda = 60$ [1/h], and the mean time between appearances of clients (passenger groups) is 1 minute.

Although the qualitative nature of the random flow of appearances is the same in every station - the value of the intensity of λ may be different for various stations, depending on their location in the network (e.g., different for the city center, for stations in the peripheral housing estates, etc.). This is a difficulty in creating a simulation model, because in the absence of actual measurement data from proposed network - designers must base on estimates. The designer may use the demographics of the area (population density) and analogies to the need for other modes of transport (buses, trams, taxis, etc.).

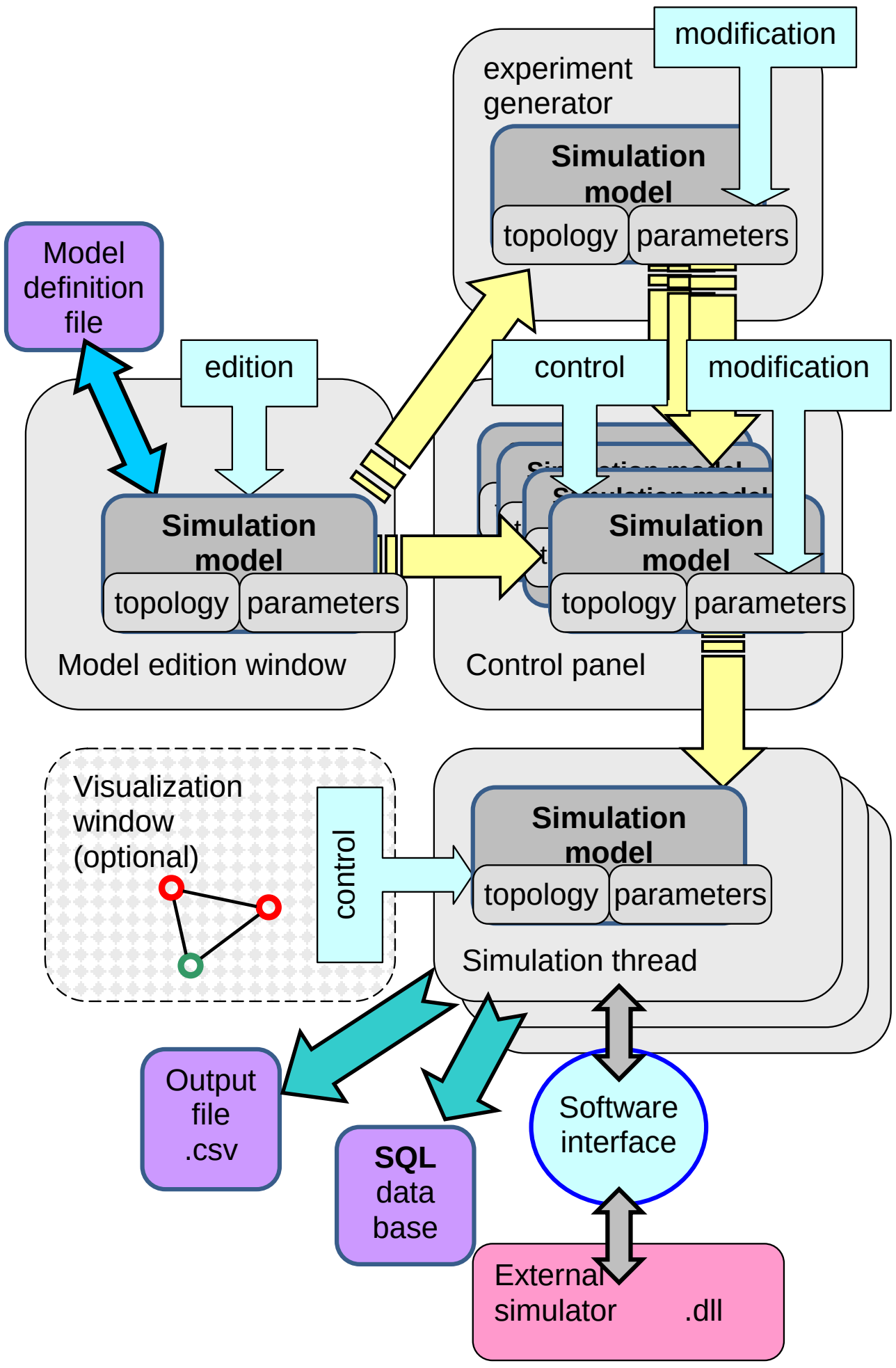

***Fig. 3** The structure of Feniks 4 simulation environment*

The second difficulty is the fact that in practice the input demand is unstable, it means that the intensity changes over time: it is different for peak hours and for off-peak times, different for weekdays and for weekends, and different in the hours before a mass social event etc. The simulation model must be able to define such a change, for example, by determining the intensity of the input for individual hours. The total intensity stream of appearances to the entire network is always the arithmetic sum of the intensities of individual stations.

It must be taken into account that this estimate may be subject to considerable error. Therefore, when creating a simulation model a safety margin should be taken, examining the behavior of the network (e.g., for the occurring of traffic jams etc.). The designer should assume the input intensity or two-three times greater than is apparent from the first estimate.

Demand model must also specify the parameters of grouping passengers undertaking a joint trip in one vehicle. In the simplest case, the cardinality of the group of passengers is random and has a value from 1 to the maximum in the model with equal probabilities.

### 1.7. Passenger Behavior

A group of passengers from the time of the appearance at the station becomes a potential customer of the PRT network. The model of the customer behavior includes the determination of the following properties:

- terms of the possible cancellation of the service,
- the distribution of boarding and disembarking times,

- how to select a destination station of the trip.

Cancellation conditions (e.g., time-out period after which the passenger group gives up and disappears from the network, or the length of the queue that they "scares") are rarely taken into account in the simulation of PRT network. Nevertheless, some simulation environments (e.g., Arena [36]) allow for such a behavior.

Time boarding and alighting of passengers is typically modeled as a random variable with triangular distribution. In the particular case, you can take the minimum time and maximum equal to the most probable, resulting in a constant time of boarding/alighting.

The selection of the destination typically uses a matrix of transport requests, known as ODM (Origin-Destination Matrix). It is a square array in which the *i*-th row and *j*-th column indicates the probability that the passenger at the station *i* selects the *j*-th station as a trip destination. The main diagonal of the matrix is composed of zeros, and the sum of the values in each row must be 1.

Similarly, as in the case of unstable input demand, the matrix ODM may vary in different periods of time, for example may be different for the morning rush, when many people go to the center, and different for the afternoon rush, when everyone goes back from the center to outlying districts, etc.

The term ODM (especially when it has to be a whole set of matrices corresponding to different times of day and different days of week) at the design stage of non-existent network requires a lot of arbitrariness. In the absence of actual statistical data a uniform matrix may be used.

## 2. DISCRETE EVENT SIMULATION

The tasks of a simulation environment (for simplicity called simulator [3,12,14,16,16,41,51]) are:

1. Define, modify, and edit the simulation model of PRT network.
2. Define the parameters of the simulation experiment (simulation time, duration of warm-up period, replications number, registering parameters: file names etc.).
3. Perform the simulation itself, combined with registration of the data generated by the simulation run.
4. The two-dimensional animation of the simulation experiment (it is optional, it can be turned on and off at any time at the user's request).
5. Processing and editing the results.

The simulator integrates the software tools related to the tasks listed above. Usually it has several modes, corresponding to individual tasks (mode of the model topography edition, defining parameters, simulation mode, animation etc.). The structure of the Feniks simulation environment is presented in Fig. 3.

The rhythm of the simulation mode of a discrete event simulator operation is defined by events attributed to PRT objects: passengers and vehicles. On the management level:

- events concerning passengers groups: the appearance of a passenger group at the station (with the determination of the destination), the seizure of the vehicle, the start and the end of trip;
- events concerning the vehicle are: seizing by a group of passengers, the start and end of the trip; these are the same events as in the case of a group of passengers (except the appearance of the group at a station), but they are considered from the point of view of the vehicle rather than of group of passengers;
- there is one additional event: the start of the trip without passengers; the "empty" trip can be initiated by the network management system of PRT.

At the coordination level the events of much finer granulation should be considered:

- for a group of passengers: taking place in the queue, leaving the queue, start and end embarking and so on;
- for a vehicle: occupation of the consecutive locations at the station area (or at the capacitor), and then moving though the sections of the track and intersections on the way to a destination, and finally reaching the target station (or capacitor) to a stop in a parking place and to disembark the passengers.

Each event is worked out to calculate all the consequences of the event, both for objects directly concerned (vehicle and/or a group of passengers), as well as for other objects in PRT network. The calculations are finished with the determining of the next event with its time stamp

At every station there is a "passenger generator", which is responsible for the creation of groups of passengers with given distribution.

All scheduled events are inserted into a structure called the event buffer, in ascending order according to their timestamps.

The simulator in the simulation loop takes the first event from the buffer, executes it (e.g., the end of the boarding of the passengers), then realizes all the consequences of this event (e.g., the end of the boarding causes the leaving a parking space by the vehicle etc.). During this procedure new events may be generated.

# 3. SIMULATION EXPERIMENTS

## 3.1. Parameters of Vehicles Control

Because the PRT vehicles are unmanned, their movement is not subject to rigid timetables and is completely automated - vehicle control algorithms are a very important part of every PRT project. The properties of these algorithms have a decisive influence on the basic indicators of network effectiveness (e.g., the number of trips per unit of time, and the average waiting time for a vehicle, average daily mileage of vehicles, etc.). Therefore simulation study of different variants of control algorithms vehicles are one of the main reasons for the use of simulation in design decisions.

There are many control algorithms in PRT network. Those that affect the behavior of the network under consideration as a transport system, can be divided into two basic classes:

- algorithms for coordinating the movement of vehicles,
- algorithms for vehicle management in the network.

The purpose of coordination algorithms is to control the behavior of a single vehicle in each of the two separate environments:

- segments of the track,
- stations and capacitors.

The behavior of the vehicle on the segments of the track includes speed control: not to exceed the maximum speed for the vehicle and for the track sector, including the restrictions on the following sectors (a vehicle should start to slow down in advance, if the next sectors maximum speed is lower). Also, specific actions as joining the traffic and coordination on “fork” intersections should be modeled.

Driving a vehicle on the track must also concert the keeping up with the preceding vehicle, if it is close enough that if may create a dangerous situation.

There are possible two general keeping up principles:

- careful principle: the vehicle can travel at most at such a speed which allows to safely stop on the track between the current position and the current position of the preceding vehicle;
- optimal rule: the vehicle can travel at a speed allowing for stopping a given distance (called static separation) behind the preceding vehicle, assuming that the preceding vehicle starts to decelerate at a time.

The optimal rule gives a smaller margin of safety (less security in the case of a failure of the vehicle or the control system), but it allows for a much larger network capacity (the number of vehicles that can simultaneously move in the network).

In modelling of a vehicle behavior such additional activities should be concerned as boarding of a group of passengers, taking a berth, taking a place in the exit buffer etc. The behavior in the capacitor is a simplified version of the behavior at the station (there is no interaction with passengers in the capacitor).

It is noteworthy that the simulation study of the various coordination algorithms may be performed on much simplified simulation models. The simulated network may consist of only a few segments of a track, one or two stations and a capacitor, two or three vehicles, and a demand model and passenger behavior may be extremely simplified. After developing a correct coordination algorithm in the case of two or three vehicles - it can be safely applied in the full model.

The simulation study of different variants of the second class of algorithms, i.e., management algorithms for a set of vehicles, should be performed on a full network model. The purpose of these algorithms is to manage a behavior of the entire large group of vehicles, which ensure the achievement of basic transport functions PRT (such as the choice of vehicle for a group of passengers), as well as the optimization of the system (for example, a management of empty vehicles to minimize queues lengths and waiting times at the stations). The management must take into account the state of a certain part of the network, but not necessarily the entire network.

For example, the problems that need to resolve dynamic management algorithms include the following questions [3,4,5,21,22,23,34,39,40,48]:

- Which vehicle is to optimally allocated for the call from the new group of passengers, appearing at the station?
- What to do with the vehicle, which has just been released? Leave it on that station, move to another station (and which) or to the capacitor? Can it be predicted where it may be needed after for a while?
- If all berths at the station are occupied, and there is an approaching vehicle with passengers, would be one of the vehicles (and which) sent somewhere (and where)?
- What route in the graph of the network will be optimal to send the vehicle (empty or full) from the starting location to the destination? Can the route be updated (optimized) during the passage, for example, depend-ing on the current density of traffic and possible congestion at various points of the network (dynamic routing)?
- … etc.

### 3.2. Routing

The last of the questions outlined above require further comment because the choice of the route is one of the essential tasks of management algorithms. In the simplest case, the network can be considered as a directed graph in which each edge is assigned a length, and one of the known algorithms for calculating the shortest path in the graph can be used (e.g., The A* algorithm, [26] or Dijkstra algorithm [27]). Directed graph of the PRT network should be strongly consistent, to reach every node from every other node. In many situations two or more alternative routes may be designated. The route should be selected optimally from a particular point of view.

The calculated route is identified by source and destination nodes, and a sequence of consecutive “fork” intersections between them (selection of the direction of travel can take place only in these intersections). The most favorable route (from a particular point of view) contains a set of segments selected in such a way that globally counted objective function reaches its minimum (of all possible routes). To calculate the value of this function, usually a number called a cost is assigned to every segment. The value of the function is a cumulative cost of all segments in the route. In the simplest case, the cost of the edge of the graph is the length of the segment, but it can also consist of multiple components (weighted with respective weights), for example, the distance, the inverse travel time, the type of the segment.

If the cost of the segment depends on the state of the track (for example, the occurrence of traffic jam or temporary failure of the segment), then the value of the objective function may vary in time. It may be useful to recalculate the route while driving the vehicle (dynamic routing).

It should be noted that since choose between two paths can only be on a crossover, it makes no sense to set the road at a moment, just before the crossover.

It should be noted that each route calculation may either increase its total length (if, for example, it is necessary to bypass the jam), as well as shorten it (if the previously rejected route improved in the meantime). Therefore, dynamic routing is a form of optimizing the travel time when the traffic conditions change.

Dynamic routing can also be used to reconfigure the network in case of failure of the track. If as the result of the failure the graph becomes inconsistent (some points become unreachable from some other ones):

- vehicles that cannot reach the end of the broken segment must simply stand up, because in PRT network the retreat is not provided;
- vehicles that cannot reach their destinations need to set a new target (probably as close as possible to the previous destination topographically);
- other vehicles should recalculate their routes, because some of the nodes in the existing routes may become unreachable (even if the destination is still achievable).

The study on various versions of vehicles management algorithms were an important part of the Eco-Mobility project and among other things, influenced the decision to implement the own PRT simulator. Available simulation environments, both commercial (Arena [36], NETSIMMOD [45]) as experimental (Hermes [35], Beamways [8], RUF [46]) usually offer their versions of management algorithms, usually already "sewn" in the simulator and practically not impossible to a modification (or very difficult to access). The simulation studies in Eco-Mobility project show that the ability to independently manipulate the parameters of algorithms for vehicle management (in particular empty vehicles management) has a very significant impact on the indicators of the PRT effectiveness.

### 3.3. Input and Output Simulation Parameters

During the simulation, the dependence of the behavior of the PRT network depending on the simulation parameters is investigated. The simulator should allow to change the values of a large number of input parameters associated with:

- the structure of the network (modification of the nodes and edges of the graph, the speed limits);
- the structure of the stations (the types of the stations, the number of the berths, the size of the buffers);
- the input and the behavior of passengers (the intensity of the input, the distribution of group sizes, the distribution times of embarkation and disembarkation);
- the origin-destination matrix;
- the vehicle movement parameters (acceleration, deceleration, speed limits, separation);
- priority rules on “join” intersections and during entering the traffic;
- routing;
- vehicle coordination and management algorithms.

Simulation environment should facilitate the organization of the simulation experiments. In particular, the important features are:

- to manage files containing models, parameters, the resulting data, meta-parameters (concerning the simulator itself rather than simulation models);
- to define a graph of nodes and segments in the model, along with the parameters of this graph (types and capacities of the stations, segment lengths, splitting into sectors, etc.); the placing of a plan of a city as a model background is the very convenient feature – the designer can determine the scale of the topographic model and apply the segment lengths straight from the plan of the city;
- to edit the graph: scaling the model, move nodes, change parameters such as the types of the stations, defining the colors of network elements;
- to change network parameters, traffic passengers, vehicles management parameters and others to generate multiple models based on the same layout of the network (with different values of the parameters of movement etc.) in a comfortable way;
- to start the simulation and manage its run: switching on/off the animation, changing the deceleration or acceleration, setting traps, moving the screen above the model (if the model does not fit on the screen), turning on/off the display of dynamic information (the numbers of passenger groups of in the queues, the numbers of vehicles in berths, etc.), interrupting the simulation.

The simulator presents the results in tables or graphs (e.g., Hermes [35]), but it can also perform these functions using publicly available programs (e.g., Microsoft Excel), especially when it concerns the compilation of results from multiple experiments.

For the convenience of defining the numerical values of the parameters, the user is often proposed the default values and he/she decides whether to use defaults or not for individual components of the model (e.g., the intensity of the input stream for all stations uniform or differing for the various stations, etc.).

The parameters of the structure of the network are constant over time simulation, but - as mentioned above - a number of parameters (of the traffic, input, management algorithms) may vary during the simulation, reflecting particular situations (e.g., variability of the input intensity and ODM matrix at different times of the day, etc.).

After the completion of the experiment, the simulator prepares - as the result of - some standard, the most commonly used output parameters, reporting the behavior of the network. The results concern:

- the stations (average queue lengths, average waiting times, average numbers of vehicles standing, etc.);
- the vehicles (average travel times, average route lengths, mileage of full and empty of trips, delays, etc.).

In addition to average values, the result of the simulation can contain:

- the report all trips (from where to where, the minimum time and the actual time, delay, the length of the route, the number of passengers)
- the report of events selected by the user (e.g., start and end of boarding, entry and exit from buffers, etc.).

These detailed reports, can be used to build the user's own, performed off-line procedures for calculating and editing simulation results.

## 4. RUN OF THE EXPERIMENT

### 4.1. Microsimulation

The simulation must be carried out with a small time step, to be able to reflect the continuous movement of vehicles, and to take into account the interactions between vehicles as accurately as possible. Modeling of the relatively small time intervals is called microsimulation [1,2,13,30,32,44,49,52]. Although some authors claim that it can be much faster and easier to perform experiments without microsimulation [31], it should be emphasized that a very limited pool of questions may be answered this way.

The basic decision about microsimulation is a method for determining the beginning of the time step, and so the moment of making decisions on a new state of the network. The solution may be synchronous, when the decisions about the behavior of vehicles are taken at the same time for all of them, with fixed time tick. The simulators based of cellular automata are built this way [9,47].

In asynchronous solution, the time step starts when a certain event occurs. The event is, for example: appearance of a new group of passengers, the end of the boarding, the arriving of a vehicle at the destination and so on. As mentioned previously, the event is also generated then any vehicle arrives at a boundary of the track sectors (small parts of the network segments). Since events may occur for individual vehicles at different time stamps, the simulation is essentially asynchronous.

The simulator running on the (asynchronous) principle is called event-driven (discrete event simulator [20,29]).

### 4.2. Simulated time flow

The simulated time should distinguished from the simulation time. For example, one hour of the simulation (in real time the run of the simulation calculations) may be subjected to simulated network within one week (the simulated time). The is no proportion between the two time flows. The simulated time runs slower or faster, depending on how the events are "handled" by the simulator. In particular, this may spoil the "smooth" animation of the PRT network.

However, the natural and the best solution is to perform the simulations with the rate determined by the work of the simulator. If the user wants to view the work of the network in a uniform rate – he/she can save a log of events from the experiment and run a separate, off-line player which displays the "smooth" presentation. Some simulation environments (e.g., Arena [36]) offer such tools.

To be closer to smooth simulation, a simulator may:

- slow down the simulation by inserting a delay loop between subsequent events during the experiment run;
- perform the simulation step by step, serving in a step or one event, ensuring the flow of a simulated time interval during predetermined period of simulation (real) time;
- provide the stopping the simulation at a specific moment of time or after a specified event.

### 4.3. Visualization and Animation

The simulator should have even a simplified graphical interface that allows editing (two-dimensional) PRT network graph, perhaps over the terrain map (or satellite photo) as the background ("model edit window" in Fig. 3).

If the simulator is used for design purposes, the animation is not actually necessary, but it is very convenient and is typically available ("visualization window" in Fig. 3). Regardless of significance for the presentation and promotion of the project results, also at the stage of simulation experiments the graphical output facilitates the evaluation of certain phenomena (e.g., the observation of the jams, starting up the simulation, etc.).

The animation is based on a dynamic visualization of the network status: the positions of all vehicles, the current lengths of the passenger queues, the number of vehicles at each station, etc. The easiest way to animate the model is based on showing the network status after each step of simulation (or after a specified number of steps). Advanced way is to simulate the process in the background, showing the visualization with a certain time step (preferably the step can be defined by the user).

Of course, the animation take much longer of the simulation time. Therefore, after running and visually checking the operation of the simulation model, at the stage of multiple simulation runs for different values of the network parameters, the animation should be switched off.

### 4.4. Warm-up Period

To gives reliable results, the simulation experiment should work in an equilibrium state. A state of equilibrium is not reached immediately, but rather after a certain time period when the vehicles ride out of capacitors and fill the whole network. If the vehicles are pre-arranged at the stations, the equilibrium state is achieved faster than in the case of initial placement of vehicles in the capacitors. In both solutions, there should be a period of time at the beginning of the simulation, in which the network is working, but data on its behavior are not collected (or are discarded later). This time is called the "warm-up" period.

Too short warm-up period (or a lack of such period) can lead to a significant distortion of simulation results.

Note that if the passenger input rate exceeds the saturation point (maximum ridership), the network never reaches the equilibrium state.

## 5. AUTOMATION OF THE EXPERIMENTS

During the research projects, the designer prepares many sets of simulation parameters, and he/she should manage many files with parameters and with results of the experiments. This work is not easy and usually is error prone. This often results in repeating sets of experiments. For example, if the designer analyzes the PRT network with constant values of all parameters exception of three parameters, and each of these three parameters is tested with four different values, the number of simulations which must be performed is 64. In addition, the long waiting for the results of multiple simulations (with multiple replications each) compromises the patience of the designer. Thus the preparation of the experiments and collecting the results should be automated ("control panel" in Fig. 3):

- the process of preparing a number of simulations in which most of the parameters have constant values, and only some of the parameters create a Cartesian product of their value sets ("experiment generator" in Fig. 3);
- the extraction of output data in order to produce tables, graphs and lists; as a repository of the results the SQL database should be applied [10,24], which allows for the preparation of compilations of the resulting using a small number of queries (or even a single query, "SQL data base" in Fig. 3);

- the distribution of simulation experiments over the parallel environments: each simulation is performed in a separate thread running in the background (main thread manages the animation, if it is enabled); multiprocessor systems allow to perform experiments actually parallel ("simulation thread" in Fig. 3).

## CONCLUSIONS

The Feniks simulation environment was developed under the Eco-Mobility project with success. The simulator follows most of the principles described in this paper. The important results of the research achieved with the help of Feniks proved that the decision to implement a customized simulation environment was correct.

The Feniks environment, in the current version, is an efficient and useful simulator, which is also confirmed by the examples of simulation studies [21,22,23]. Especially worth mentioning is the development of multi-parameter algorithms: vehicle management and dynamic route selection. They enable the creation and testing of a wide variety of strategic vehicle management options in the PRT network. Another important feature is the ability to run multiple simulation runs at the same time, in parallel processing mode, which greatly improves and accelerates research. Finally, a good solution should be considered to record the simulation results in a database managed by SQL server. This simplifies the processing of the results in the various sections of data.

Thanks to all of the above properties, the Feniks simulator is advantageous over other solutions known from the literature. Further development of this simulation environment should assume improvement of the management algorithms, further improving edition mechanisms of simulation models, as well as supporting the analysis of specific case studies, based on the actual demographic data and the topography of the terrain.

A large number of trials and complete simulation experiments also led to the elaboration of a methodology for simulation experiments with models of PRT network. In particular, the experiments allowed to elaborate relatively realistic estimation of the demand for PRT network services, to find proper number of vehicles for a given intensity of the input stream, to get the maximum ridership and safety margins, etc. Also, a set of four benchmark models was prepared, corresponding to a typical urban layouts. Simulation of network behavior in a number of different benchmarks and comparison of results, allow to draw conclusions more general than is the case of the specific case studies, referring to the very specific conditions of a particular PRT network.

The local administration authority, the administrator of the commercial area, airport, etc. should be aware of the advantages of PRT network and its technical feasibility with the current state of the art. Of course, its possible construction must in any case be based on a deeper analysis of the local conditions, taking into consideration the local aspects of financial, legal and urban planning. Simulation of the PRT network using the Feniks environment can be a tool, providing data and documented estimates necessary for such an analysis.

## BIBLIOGRAPHY

1. Akçelik R., Besley M., Microsimulation and analytical methods for modelling urban traffic. *Proc. Conference on Advance Modeling Techniques and Quality of Service in Highway Capacity Analysis*, Truckee, California, July 2001
2. Algers S., Bernauer E., Boero M., Breheret L., Taranto C.D., Dougherty M., Fox K., Gabard J.F., *Review of Micro-Simulation Models*. SMARTEST (Simulation Modelling Applied to Road Transport European Scheme Tests) Project Report. Institute of Transport Studies, University of Leeds, UK, 2000
3. Anderson J.E., Simulation of the operation of personal rapid transit systems. in *Computers in Railways VI*, WIT Press, Boston Southampton, Computational Mechanics Publications, 1998, pp.523-532
4. Andréasson I.J., Vehicle distribution in large personal rapid transit systems, *Transportation Research Record 1451*, 1994, pp.95-99
5. Andréasson I.J., Quasi-optimum redistribution of empty PRT vehicles, *Proc. 6th International Conference on Automated People Movers*. Las Vegas, Nevada, April 9-12, 1997, pp.541-550
6. Andréasson I.J., *Reallocation of Empty PRT vehicles en route*. TRB annual meeting, Washington DC January 2003
7. Bazzan A.L.C., de Brito do Amarante M., Da Costa F.B.L., Management of Demand and Routing in Autonomous Personal Transit. *Journal of Intelligent Transportation Systems: Technology, Planning, and Operations* Volume 16, Issue 1, 2012, pp.1-11, DOI: http://dx.doi.org/10.1080/15472450.2012.639635
8. Beamways, http://www.beamways.com/
9. Benjamin S.C., Johnson N.F., Hui P.M., Cellular automata models of traffic flow along a highway containing a junction, *Journal of Physics A: Mathematical and General* 29 (12) (1996) pp.3119–3127, DOI: http://dx.doi.org/10.1088/0305-4470/29/12/018
10. Bowman J.S., Emerson S.L., Darnovsky M., *The practical SQL handbook: using structured query language*, 3th ed., Addison-Wesley 1996, Readin, MA,  ISBN:0-201-44787-8

11. Cahill V., Senart A., Schmidt D.C., Weber S., Harrington A., Hughes B., The Managed Motorway: Real-time Vehicle Scheduling – A Research Agenda, *Proc. HotMobile 2008, 9th workshop on Mobile computing systems and applications*, ACM, New York, NY, USA, ISBN: 978-1-60558-118-7, pp.43-48, DOI: http://dx.doi.org/10.1145/1411759.1411771
12. Carnegie J.A., Hoffman P.S., *Viability of Personal Rapid Transit In New Jersey* (Final Report). Presented to Governor Jon S. Corzine and the New Jersey State Legislature. February 2007, http://policy.rutgers.edu/vtc/reports/REPORTS/PRTfinalreport.pdf
13. Casas B.J., Ferrer J.L., García D., Modelling Advanced Transport Telematic Applications with Microscopic Simulators: the Case of AIMSUN2. *Proc. Micro-simulation workshop*, Institute for Transport Studies, University of Leeds , July 1999, pp.11-24, DOI: http://dx.doi.org/10.1007/978-3-642-60236-8_14
14. Castangia M., Guala L., Modelling and simulation of a PRT network. *Proc. 17th International Conference on Urban Transport and the Environment*, June 6-8, 2011, Pisa, Italy, pp.459-472
15. Choromański W., Daszczuk W.B., Grabski W., Dyduch J., Maciejewski M., Brach, P., Personal Rapid Transit (PRT) Computer Network Simulation and Analysis of Flow Capacity. *Proc. 14th International Conference on Automated People Movers and Transit Systems* 2013:, April 21-24, 2013, Phoenix, pp. 296-312, DOI: http://dx.doi.org/10.1061/9780784412862.022
16. Choromański W., Daszczuk W.B., Dyduch J., Maciejewski M., Brach P., Grabski W., PRT (Personal Rapid Transit) Network Simulation, *Proc. 13th World Conference on Transportation Research*, Joao Victor (ed.), 2014, Federal University of Rio de Janeiro, ISBN 978-85-285-0232-9
17. Choromański W., Kowara J., PRT – modeling and dynamics simulation of track and vehicle. *Proc. 13th International Conference on Automated People Movers and Automated Transit Systems*, Paris, France, 22-25 May 2011, pp. 294-306, DOI: http://dx.doi.org/10.1061/41193(424)28
18. Choromański W., Kowara J., Personal rapid transit vehicle with polyurethane wheels – modelling and simulation issues. *Archives of Transport*, Vol. 27-28, Issue 3-4, 2013, pp. 71-79
19. Choromański W., Kozłowski M., Kowara J., Analysis of Dynamic Properties of the PRT Vehicle-Track System. *Bulletin of the Polish Academy of Sciences, Technical Sciences*, Volume 63, Issue 3, September 2015, ISSN (Online) 2300-1917, pp.799–806, DOI: http://dx.doi.org/10.1515/bpasts-2015-0091
20. Dagpunar J.S., *Simulation and Monte Carlo*, 2007 John Wiley & Sons, Ltd, ISBN: 9780470854945, Chapter 7. Discrete Event Simulation
21. Daszczuk W.B., Choromański W., Mieścicki J., Grabski W., Empty vehicles management as a method for reducing passenger waiting time in Personal Rapid Transit networks. *IET Intelligent Transport Systems*, vol. 9(2015), No.3, pp. 231 – 239, DOI: http://dx.doi.org/10.1049/iet-its.2013.0084
22. Daszczuk W.B., Mieścicki J., Distributed management of Personal Rapid Transit (PRT) vehicles under unusual transport conditions. *Proc. XII LogiTrans conference*, Szczyrk, 20-23.04.2015, *Logistyka* Vol. 4/2015, pp. 2896-2901.
23. Daszczuk W.B., Mieścicki J., Grabski W., Distributed algorithm for empty vehicles management in personal rapid transit (PRT) network. *Journal of Advanced Transportation*, on-line early view, 9 February 2016, DOI: http://dx.doi.org/10.1002/atr.1365
24. Date C.J. and Darwen H., *A Guide to SQL Standard*. Addison-Wesley 1994, Reading, MA, ISBN 020155822X
25. Deflorio F.P., Simulation of requests in demand responsive transport systems. *IET Intelligent Transport Systems*, Vol. 5, Issue 3, 2011, pp. 159–167, DOI: http://dx.doi.org/10.1049/iet-its.2010.0026
26. Delling D., Sanders P., Schultes D., Wagner, D., Engineering route planning algorithms. In *Algorithmics of large and complex networks*. Springer, 2009, ISBN: 978-3-642-02093-3, pp. 117–139, DOI: http://dx.doi.org/10.1007/978-3-642-02094-0_7
27. Dijkstra E.W., A note on two problems in connexion with graphs. *Numerische Mathematik* 1, 1959, pp. 269–271, DOI: http://dx.doi.org/10.1007/BF01386390
28. Eco-Mobility, http://www.eco-mobilnosc.pw.edu.pl/?sLang=en
29. Fishman G.S., *Principles of discrete event simulation*, John Wiley and Sons, New York, NY, 2011, ISBN:0471043958
30. Fox K., Introduction to the SMARTEST Project. *Proc. Micro-simulation workshop*, Institute for Transport Studies, University of Leeds, July 1999, pp.2-10
31. Fox K., Is Micro-Simulation a Waste of Time? *Proc. European Transport Conference, Proc. Networks and traffics assignment*, 6-8 October 2008, Noordwijkerhout, The Netherlands
32. Gabard J.F., Breheret L., The SITRA-B+ Microscopic Traffic Simulation Model - Examples of Use and Future Developments. *Proc. Micro-simulation workshop, Institute for Transport Studies*, University of Leeds, July 1999, p.49-58
33. Gross D., Shortle J.F., Thompson J.M., Harris C.M., *Fundamentals of Queueing Theory*, Wiley Series in Probability and Statistics, 2008, ISBN: 978-0-471-79127-0

34. Gustafsson J., Kang J.-G., Englund J., Grimtell P., Design Considerations for Capacity in PRT Networks. *Proc. Automated People Movers and Transit Systems* 2011, pp. 385-394, DOI: http://dx.doi.org/10.1061/41193(424)35
35. Hermes, http://students.ceid.upatras.gr/~xithalis/simulation_en.html
36. Kelton W.D., Sadowski R.P., Sturrock D.T., *Simulation with Arena*, 5th ed., McGraw Hill Book Co., 2010, ISBN-10: 0072919817
37. Kleinrock L., *Queueing Systems. Volume 1: Theory*, Wiley Interscience, 1975. ISBN 0-471-49110-1
38. Kozłowski M, Choromański W, Kowara J., Parametric sensitivity analysis of ATN-PRT vehicle (automated transit network-personal rapid transit). *Journal of Vibroengineering*, Vol. 17 Issue 3, May 2015, ISSN 1392-8716, pp.1436-1451
39. Lees-Miller J.D. and Wilson R.E., Sampling for Personal Rapid Transit Empty Vehicle Redistribution, *Proc., 90th Annual Meeting of the Transportation Research Board*, Vol. 2216, (2011), pp.174-1812011, DOI: http://dx.doi.org/10.3141/2216-19
40. Lees-Miller J.D. and Wilson R.E., Proactive empty vehicle redistribution for personal rapid transit and taxis, *Transportation Planning and Technology*, 35(1), 2012, ISSN 0308-1060, pp.17–30, DOI: http://dx.doi.org/10.1080/03081060.2012.635414
41. Lopes C.V., Lindstrom C., Virtual Cities in Urban Planning: The Uppsala Case Study, *Journal of Theoretical and Applied Electronic Commerce Research* Vol.7, Issue 3, December 2012, pp.88-100, DOI: http://dx.doi.org/10.4067/S0718-18762012000300009
42. Melki A., Hammadi S., Dynamic Management of Intelligent Urban Vehicles. *Proc. 11th International IEEE Conference on Intelligent Transportation Systems*, Beijing, China, October 12-15, 2008, pp.1095-1100, DOI: http://dx.doi.org/10.1109/ITSC.2008.4732712
43. Mieścicki J., Daszczuk W.B., Proposed benchmarks for PRT networks simulation, *Archives of Transport*, vol. 27-28, No.3-4, 2013, pp.123-133
44. Millard M., Evaluation of The Benefits of Active Traffic Management Schemes Using Microsimulation Programming, *Proc. European Transport Conference*, 2009, http://abstracts.aetransport.org/paper/download/id/3138
45. NETSIMMOD, http://prtconsulting.com/simulation.html
46. RUF, http://www.ruf.dk/
47. Schadschneider A., Schreckenberg M., Cellular automaton models and traffic flow, *Journal of Physics A: Mathematical and General* 26 (1993), pp.679–683
48. Schweizer J., Danesi A., Mantecchini L., Traversi E., Caprara A., Towards a PRT capacity manual. *PRT@LHR Conference Proceedings* (21-23 September 2010)
49. Szillat M.T., *A Low-level PRT Microsimulation*. PhD thesis, University of Bristol, April 2001, DOI: http://dx.doi.org/10.1.1.112.1156
50. Wang F.; Yang M.; Yang., Dynamic Fleet Management for Cybercars, *Proc. IEEE ITSC 2006 IEEE Intelligent Transportation Systems Conference*, Toronto, Canada, September 17-20, 2006, ISBN: 1-4244-0093-7, pp.1246-1250, DOI: http://dx.doi.org/10.1109/ITSC.2006.1707393
51. Xu J., Abdulrab H., Itmi M., A multi-agent based model for urban demand-responsive passenger transport services. *Proc. 2008 International Joint Conference on Neural Networks* (IJCNN 2008), DOI: http://dx.doi.org/10.1016/j.eswa.2014.05.015
52. Yand M., Macro versus Micro Simulation Modeling Tools. *ITE District 6 Annual Meeting Proceedings*, DKS Associates, 2007
53. Zheng, P., Jeffery, D. and McDonald, M., Development and Evaluation of Traffic Management Strategies for Personal Rapid Transit, *Proc. Industrial Simulation Conference* 2009, Loughborough, UK, 1-3 Jun 2009, pp.191-195, DOI: http://dx.doi.org/10.1.1.368.8499

Author:

**Wiktor B. Daszczuk**, PhD – Warsaw University of Technology, Institute of Computer Science, wbd@ii.pw.edu.pl